\documentclass[%
reprint,
 amsmath,amssymb,
 aps,
]{revtex4-1}

\usepackage{graphicx}

\begin{document}

\title{True Differential Superconducting On-Chip Output Amplifier}

\author{Jonathan Egan, Andrew Brownfield and Quentin Herr}
\affiliation{Northrop Grumman Corp., Baltimore, MD 21240}

\thanks{This research is based upon work supported in part by the
  ODNI, IARPA, via ARO. The views and conclusions contained herein are
  those of the authors and should not be interpreted as necessarily
  representing the official policies or endorsements, either expressed
  or implied, of the ODNI, IARPA, or the U.S. Government.}

\date{14 April 2021}

\begin{abstract}
The true-differential superconductor on-chip amplifier has
complementary outputs that float with respect to chip ground. This
improves signal integrity and compatibility with the receiving
semiconductor stage. Both source-terminated and non-source-terminated
designs producing 4\,mV demonstrated rejection of a large common mode
interference in the package. Measured margins are $\pm$8.5\% on the
output bias, and $\pm$28\% on AC clock amplitude. Waveforms and eye
diagrams are taken at 2.9-10\,Gb/s. Direct measurement of bit-error
rates are better than the resolution limit of 1e-12 at 2.9\,Gb/s, and
better than 1e-9 at 10\,Gb/s.
\end{abstract}

\maketitle

On-chip output amplifiers convert the aJ signal-energy of
superconductor single-flux-quantum (SFQ) logic to levels that can be
detected by standard electronics. A design using a series-array of
inductively-isolated, DC-biased SQUIDs has proven to be uniquely
effective \cite{herr2007inductive}. The design is amenable to
distributed amplifier techniques for improved gain-bandwidth product
\cite{herr2010high}. An output voltage of 2\,mV is adequate for direct
interface to a high-performance room-temperature LNA at a data rate of
10\,Gb/s \cite{herr2010high}. These findings have recently been
verified independently \cite{takeuchi2017measurement,
  gupta2019digital}. The only alternative is the SFQ-to-dc converter
\cite{likharev1991rsfq}, a smaller design that produces only
0.5\,mV. This is effective at modest data rates, but 1-10\,GHz rates
require multistage amplification at intermediate temperature stages
\cite{gupta2019digital}.

Reciprocal Quantum Logic (RQL) is an SFQ logic family that is uniquely
attractive in terms of power dissipation and latency
\cite{herr2011ultra}. RQL is AC powered, which adds a significant
constraint to the output link. The AC power delivered at the clock
frequency is typically 45-50\,dB larger than the output signals. These
power levels produce strong interference in the pressure-contact
package, where isolation is imperfect. As the output amplifiers have
narrow operating margins on the applied dc bias current, interference
large enough to shift the bias and will not only obscure the output
signal but prevent it from being generated in the first place.

The solution is to make the driver differential. This is common
practice in high-speed interconnect due to its rejection of
common-mode interference and noise. Transistor-based designs typically
produce the differential signals with two independent drivers that
contact chip ground. Amplitude and delay must be well-matched, which
presents a design challenge. The stacked-SQUID design can be
true-differential, meaning that the outputs are floating with respect
to chip ground.  This is accomplished by wiring out both ends of the
series array. The true-differential design rejects common-mode
interference and noise on both the output signal and on the bias
current.

A 2\,mV output amplifier has been used routinely for lab
demonstrations up to 3.5\,Gb/s using high performance LNAs on the
receiving end at room temperature. The circuit converts the SFQ-based
data encoding of RQL to return-to-zero (RZ) encoding. An important
feature of the output amplifiers described here is conversion to
standard non-return-to-zero (NRZ) encoding, which has twice the
bandwidth efficiency, enabling 10\,Gb/s data rates. These designs
produce 4\,mV. A future target is direct drive of a differential
limiting amplifier (e.g., Hittite HMC750LP4) with improved size,
weight, and power (SWAP). We report measurements from a sampling of
parts, including 10\,GHz operation of a circuit fabricated internally
at Northrop Grumman in a Nb-based six-metal-layer integrated circuit
process.

\section{Design}

\begin{figure}
 \centering
 \includegraphics[width=3.45in]{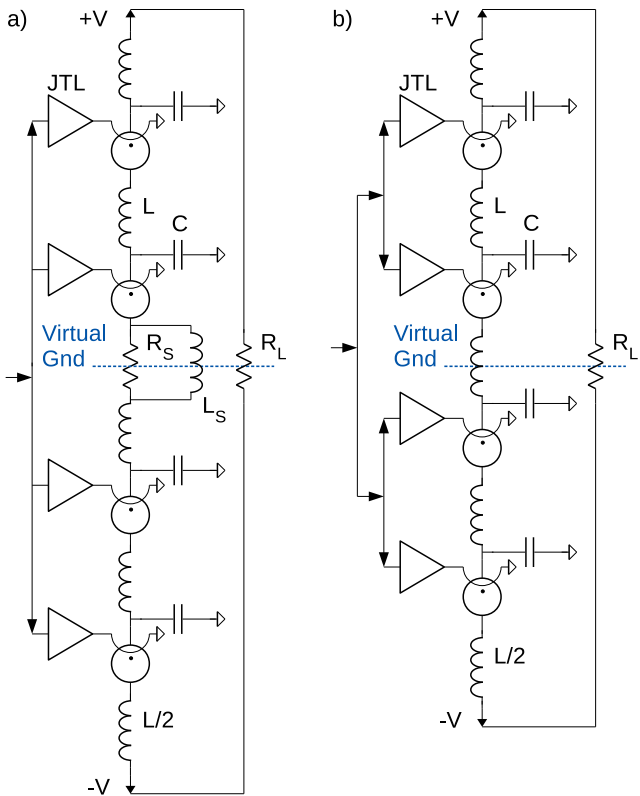}
 \caption{True-differential output amplifiers driving load resistor
   $R_L$. Both designs use a series-array of SQUID-based floating
   voltage sources embedded in an $LC$ matching network. a) The
   source-terminated design has the resistor $R_S=R_L$ in the center
   of the array. Half of the signal voltage is dissipated across this
   resistor. Bias current flows through the optional bypass inductor
   $L_S$. b) Without source termination, the design generates the same
   voltage with only half as many stages. Both circuits have identical
   JTL networks on either side of the stack to produce an NRZ signal. The
   second JTL network is omitted from the figure for clarity.
  \label{stack}}
\end{figure}

The primary design trade-off is between source-terminated and
non-source-terminated. The non-source-terminated circuit is half the
size and twice as efficient. However, source termination is required
to scale to the higher power-gain product of distributed amplifiers,
and to terminate the large signal reflections at the receiving end
typical of low-noise LNAs.

As shown in Fig.~\ref{stack}, both designs consist of a series array
of SQUIDs connected by an $LC$ network, as with the previously
reported distributed amplifier \cite{herr2010high}. The stack is
DC-biased with current flowing from the +V to -V nodes. Each SQUID is
a low-impedance, floating voltage source producing a sub-mV
level. These voltages sum at the output. The lumped inductance and
capacitance at every stage produces a filter matched in frequency (and
impedance) to the output waveform, and mismatched to the internal
oscillatory modes in the SQUID and between multiple SQUIDs. This is
the essential function of inductive isolation
\cite{herr2007inductive}.

The source-terminated design is effectively two single-ended
distributed amplifiers placed back-to-back. Signal propagation time of
$\sqrt{LC}$ per stage produces a distributed design that exceeds the
bandwidth-gain product of the individual stages. By timing the trigger
for each stage, the total rise time can be as small as that of a
single stage. The differential signals generated in the array or
reflected off the load are equal in amplitude and of opposite polarity
as seen by the source termination. This produces a virtual ground,
without explicit grounding. The bypass inductor eliminates power
dissipation of the bias current. To prevent distortion, the $L_S/R_S$
time constant must be large enough to outlast DC offsets that occur in
the data pattern, e.g. in a long series of ``all-1’s'' or ``all-0’s''.

The non-source-terminated design produces the same output voltage with
half the number of stages. This design also has a virtual ground
across the central inductor, but here differential signals will be
reflected, not terminated.  Without source termination, the design is
not amenable to the sequential timing of the distributed
amplifier. Instead, the fastest risetime comes by triggering the stack
broadside. While this limits the gain-bandwidth product, the design is
efficient for modest data rates and output amplitudes.

\begin{figure}
 \centering
 \includegraphics[width=3.5in]{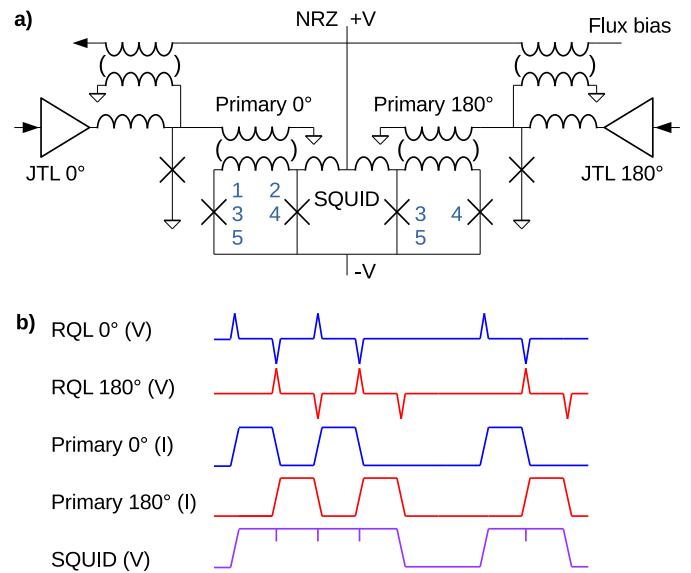}
 \caption{SQUID design. a) The four-junction SQUID has two independent
   input transformers. The numbers label the switching order of the
   Josephson junctions when the current goes high in the transformer
   on the left. b) SFQ pulses are applied using RQL data encoding from
   the left (blue) and right JTLs (red) with a relative delay of
   180$^{\circ}$, or half a clock cycle. The integral of the voltage
   transients is superconductor phase, equal to current through the
   primary. The resulting output voltage (purple) is a sum of the two
   phase states. The transient glitches in the output waveform at the
   input transitions are out-of-band of the output.
  \label{squid}}
\end{figure}

Here we report two circuits designed to produce 4\,mV with 50\,ps
risetime, supporting data rates of 10\,Gb/s. Both circuits
convert RQL data encoding to standard NRZ-encoded data at the output,
with a fundamental frequency that is half the data rate. As shown in
Fig.~\ref{squid}, the conversion is done using a four-junction SQUID
that couples signals in through two independent transformers. The RQL
pattern applied to the left is delayed by half a cycle and reapplied
on the right. The sum of these two return-to-zero (RZ) voltage
waveforms produces an NRZ waveform. The design uses two separate JTL
networks to feed the output. The feeding JTL and the SQUID are
isolated from each other using an unpowered junction. This isolation
reduces AC power leakage from JTLs to the SQUID, and reduces
back-action from a selected SQUID to the JTLs. A flux bias produces
the desired symmetric double-well potential so that the unpowered
junction can be triggered.

Advantages of the four-junction SQUID over a conventional two-junction
design include: 1) Smaller junctions and larger inductors in the SQUID
produce more efficient transformer coupling. These designs did not use
holes in the ground plane under the transformers in physical
layout. 2) More junctions in the SQUID improves current
compliance. Total critical current of 400\,$\mu$A produced the
necessary 40\,$\mu$A output signal swing, corresponding to 4\,mV into
the 100\,$\Omega$ differential load.

The disadvantage of the four-junction SQUID is relatively modest
voltage, 0.25-0.30\,mV per stage using Josephson junctions with
critical current density of 100\,$\mu$A/$\mu$m$^2$. With two inputs,
the SQUID is only ever partially selected. As illustrated in
Fig.~\ref{squid} by the start-up switching order of the junctions, the
selected side of the compound SQUID effectively drives fanout to the
other side.

\begin{table}[b]
\renewcommand{\arraystretch}{1.0}
\caption{Parameters of RQL output amplifiers}
\label{power}
\centering
\begin{tabular}{cl|rr|rrr}
  \multicolumn{1}{l}{{\bf Output}} &
  \multicolumn{1}{l}{{\bf Source}} &
  \multicolumn{1}{|l}{{\bf SQUID}} &
  \multicolumn{1}{l|}{{\bf Area}} &
  \multicolumn{1}{l}{{\bf Power}} &
  \multicolumn{1}{l}{{\bf Output}} &
  \multicolumn{1}{l}{{\bf Effcy}} \\
  (mV) & {\bf Term} & ($\times$) & ($\mu$m$^2$) & (nW) & (nW) & (\%) \\
\hline
  2$^*$   & No  &  8 &   8k &   40 &  10 & 20 \\
  4\,\,\, & No  & 16 &  61k &  160 &  80 & 33 \\
  4\,\,\, & Yes & 32 & 190k &  320 &  80 & 16 \\
  8\,\,\, & Yes & 64 & 240k & 1280 & 320 & 16 \\
\hline
  \multicolumn{7}{l}{$^*$This amplifier produces RZ output. All others, NRZ.} \\
\end{tabular}
\end{table}

Only the two junctions on the ends of the four-junction SQUID are
resistively shunted. As reported earlier \cite{herr2007inductive,
  herr2010high} such creative damping increases the output voltage and
the reliability of the transition back to zero. Total damping is
$I_cR_s=1.2$\,mV where $R_s$ is the local parallel shunt resistance
and $I_c$ is the total critical current of the SQUID. This is well
above the critical damping value of 0.7\,mV. However, the whole stack
is also shunted by load resistance and source termination resistors,
with a total effective damping of $I_cR_s=0.8$\,mV, This approximates
critical damping. Simulated margins on the dc bias of the SQUID stack
are about $\pm$10\%.

The 4\,mV design balances speed, output amplitude, and chip
area. Parameters of the 4\,mV design, in comparison to 2\,mV and
8\,mV, are given in Table~\ref{power}. RQL output amplifiers are quite
efficient with on-chip power dissipation on a nW-scale, which is
negligible compared to the heat load of the cabling. On-chip area of
61k\,$\mu$m$^2$ is dominated by that JTL feed network, and is subject
to further miniaturization in advanced fabrication processes. Even the
largest design is about the same area as that of the associated signal
and ground pads in the current package.

\section{Measurements}

All test circuits had a differential input transformer and JTL chain
\cite{herr2011ultra} connected to the output amplifier. Circuits were
powered using a resonant network that provisioned the active area of
each chip \cite{talanov2017clock}.

True-differential output amplifiers have been used on nearly all RQL
digital designs, so various configurations, clock frequencies, and
fabrication processes have been exercised. The circuits presented here
were fabricated either in the process supplied by D-Wave
\cite{berkley2010scalable}, or using the ``RQL25'' process developed
internally at Northrop Grumman. Both processes have six metal layers
and Josephson junctions with 100\,$\mu$A/$\mu$m$^2$ critical current
density, and support similar layout style with 0.25\,$\mu$m minimum
feature size.

\begin{figure}
 \centering
 \includegraphics[width=3.5in]{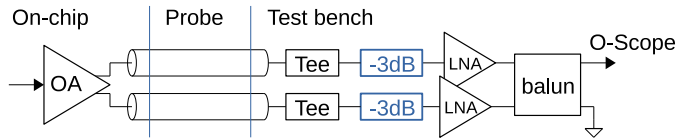}
 \caption{Output channel. Differential outputs are connected
   to micro-coax cables via pressure contacts in a liquid He dip
   probe. DC bias is delivered to the output SQUIDs via the
   bias-Tees. The 3\,dB attenuators (50\,$\Omega$) improve the
   impedance match, and are used to test non-source terminated
   amplifiers. Signals are amplified by Miteq JSMF4-02K180-30-10P LNAs
   at the probe head. The LNAs fed a Marki BAL-0026 balun to
   produce signal-ended signals for the lab instruments.
  \label{bench}}
\end{figure}

\begin{figure}
 \centering
 \includegraphics[width=3.7in]{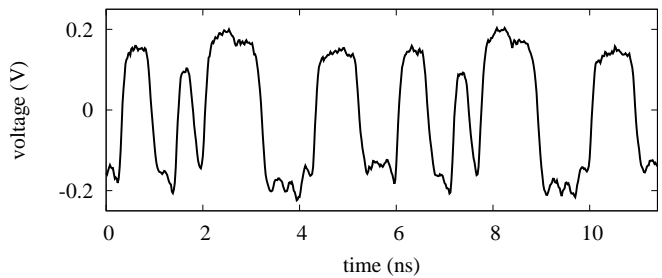}
 \caption{Measured waveform at 3.5\,Gb/s of the source-terminated
   circuit fabricated at D-Wave. Transition rates are limited to
   320\,ps by a low-pass filter added to the output channel, not by
   the part itself. In this and subsequent figures, the waveform is a
   signal-averaged ``4-3-2-1'' chirp pattern (repetitive
   11110000111000110010), and the amplitude is that seen at the
   instrument after arbitrary levels of amplification.
  \label{waveST}}
\end{figure}

\begin{figure}
  \centering
 \includegraphics[width=3.7in]{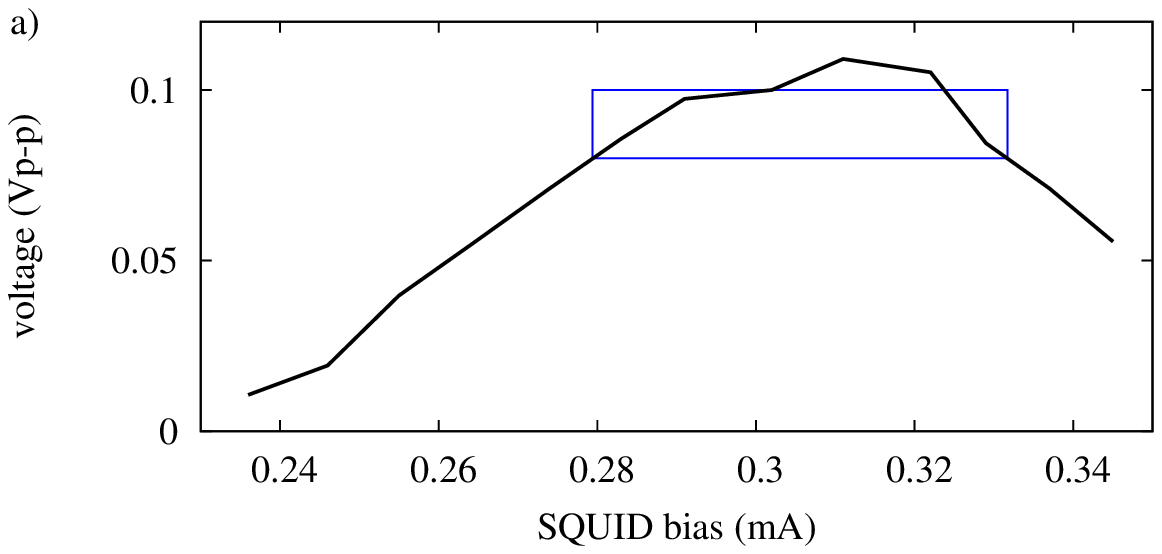}
 \includegraphics[width=3.7in]{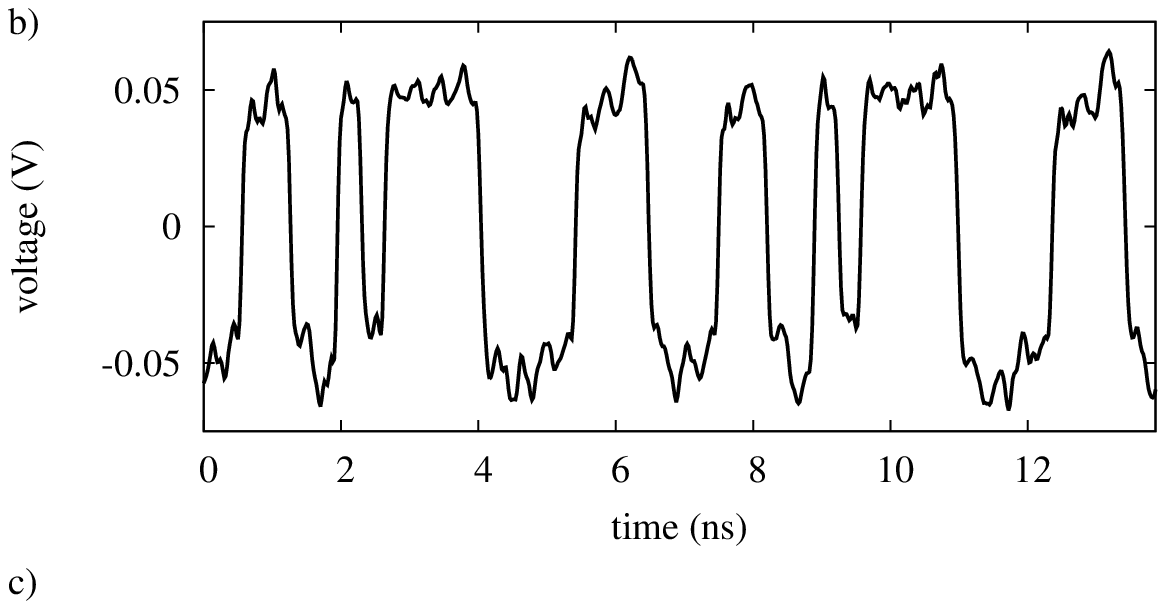}
  \includegraphics[width=3.5in]{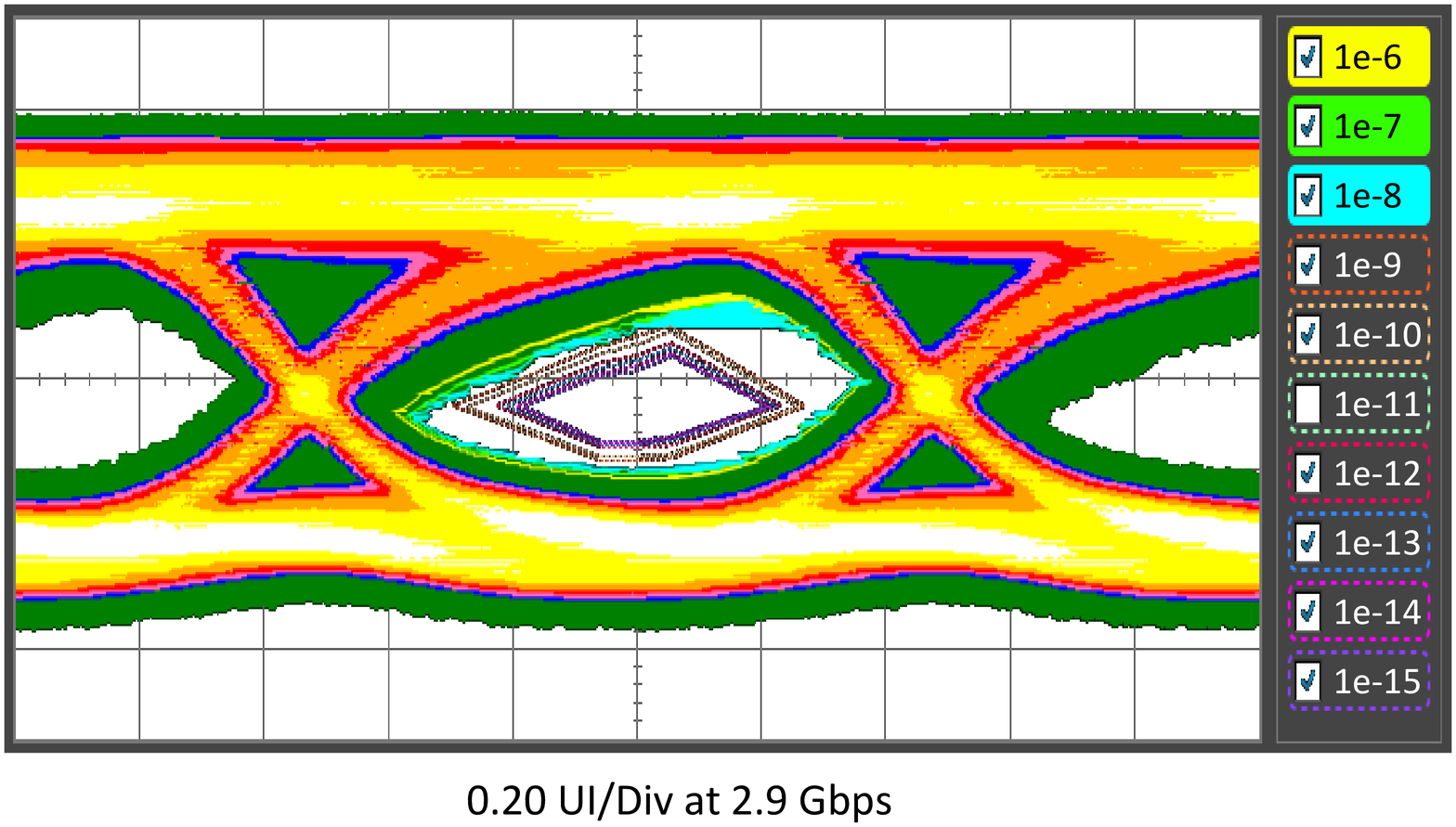}
  \caption{Measurement at 2.9\,Gb/s of the non-source-terminated
    circuit fabricated in RQL25. a) Output amplitude as a function of
    bias current. The rectangle corresponds to bias excursions of
    $\pm$8.5\% that produce output amplitude 20\% less than
    nominal. b) The waveform is the signal-averaged 4-3-2-1 chirp
    pattern defined in the previous figure. c) An eye diagram measured
    for a PRBS15 pattern. Data eye analysis on the Keysight J-BERT
    8020A shows an open eye directly measured down to a BER of 1e-8,
    and extrapolated down to 1e-15 at the inner contour.
  \label{wave2_9}}
\end{figure}

The output amplifiers were tested using the channel shown in
Fig.~\ref{bench}. The room-temperature LNAs used to amplify the
on-chip signal have a typical VSWR specification of 2:1, meaning up to
half the signal amplitude is reflected. For the non-source-terminated
circuits, $-$3\,dB attenuators were used to improve the impedance
match at the expense of cutting the signal-to-noise ratio in
half. These attenuators were not used for the source-terminated
circuits.

\begin{figure}
  \centering
  \includegraphics[width=3.7in]{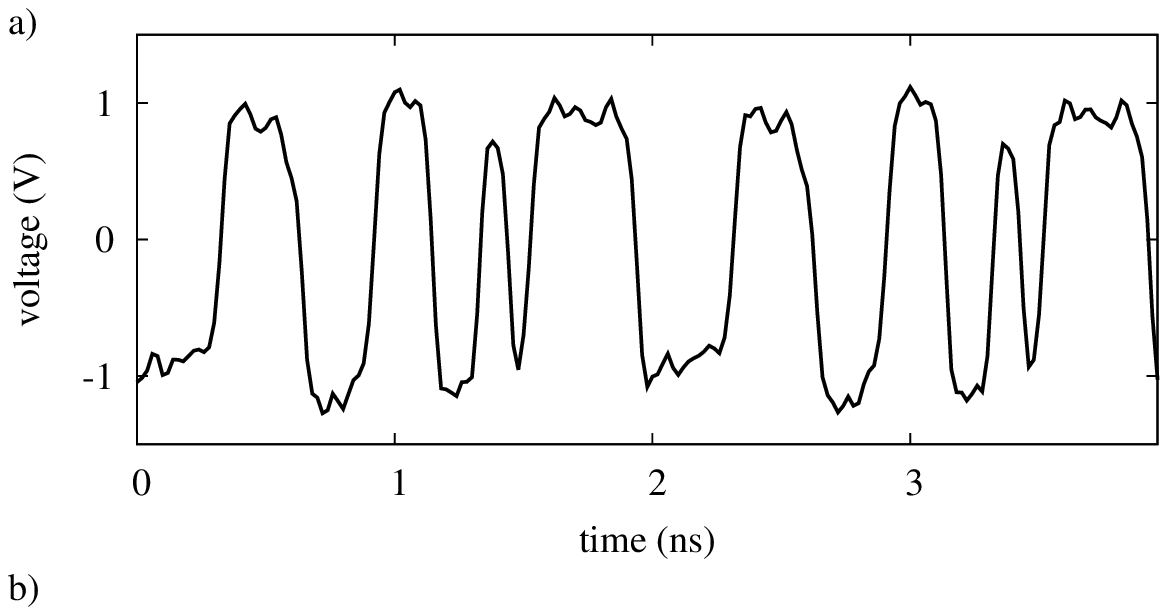}
  \includegraphics[width=3.5in]{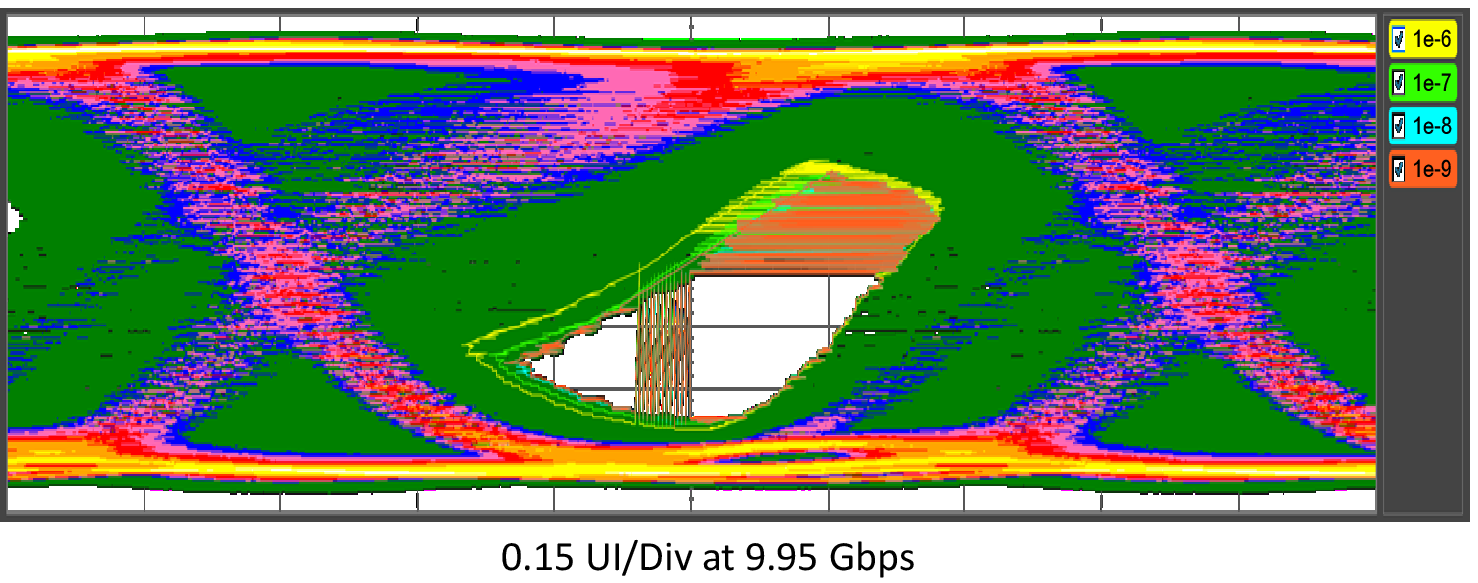}
  \caption{Measurement at 9.95\,Gb/s of the non-source-terminated
    circuit fabricated in RQL25.  a) The waveform is again the
    signal-averaged 4-3-2-1 chirp pattern. b) An eye diagram measured
    for a PRBS15 pattern. Data eye analysis on the Keysight J-BERT
    8020A shows an open eye directly measured down to a BER of
    1e-9. The secondary rising edge in the eye corresponds to the
    isolated ``101'' token.
  \label{wave995}}
\end{figure}

Here we present measurements of waveforms, eye diagrams, and
bit-error-rates (BER). Waveform captures are useful as an initial test
of signal integrity, as clock and data feed-through, and signal
reflections are visible by inspection. Eye diagrams are useful to
characterize bandwidth and signal-to-noise of the output data link as
a whole. Direct measurements of BER are required to detect switching
errors in the output amplifier. This is particularly motivated by the
history of elevated switching errors associated with superconducting
latching logic and output amplifiers \cite{fulton1971switching,
  suzuki1988josephson}. Underlying each voltage waveform is a large,
integral number of SFQ pulses, and sometimes the circuit cannot
``decide'' exactly how many pulses to produce. This is a manifestation
of Buridan's Principle \cite{lamport2012buridan}.

Fig.~\ref{waveST} shows the measured waveform of a 4\,mV
source-terminated output amplifier at 3.5\,Gb/s. The data show good
signal integrity without clock feed-through, confirming effectiveness
of the true-differential design. This compares quite favorably to
previous results with a single-ended output, where the clock
feed-through was as large as the signal \cite{herr20138}.

Fig.~\ref{wave2_9} shows the measured output amplitude, waveform and
eye diagram of the 4\,mV non-source-terminated circuit at 2.9\,Gb/s.
Output amplitude depends on the DC bias current applied to the SQUID
stack. The peak-to-peak voltage after amplification is 0.10\,V at the
nominal bias of 0.30\,mA. Bias margins of $\pm$8.5\% are measured for
output voltage within $\pm$20\% of nominal. Margins on the AC clock
that powers the JTL feed network were measured to be 5.0\,dB, or
$\pm$28\% in amplitude. This agrees with simulation. The waveform
shows good signal integrity, without clock feed-through. The eye
diagram at 2.9\,Gb/s shows good performance, as BER is directly
measured down to 1e-8 and extrapolated down to 1e-15. Additional
direct measurements of BER using the PRBS15 pattern found the data
link to be error-free down to the resolution limit, 1e-12.

Fig.~\ref{wave995} shows the measured waveform and eye diagram of the
same 4\,mV non-source-terminated design, but now clocked at
9.95\,Gb/s. The waveform again shows good integrity without clock
feed-through, but shows reduced amplitude for the standalone ``101''
token, attributable to an observed risetime of 60\,ps, which is a
little slower than the 50\,ps design value. This token is visible as a
secondary rising waveform in the eye diagram. The amplified waveform
is clipped by the instrument, as the 2\,V peak-to-peak input has an
apparent amplitude of about 1.6\,V peak-to-peak in the eye
diagram. Despite this test artifact, BER was directly measured down to
1e-9.

\section{Conclusion}
The 4\,mV true-differential output amplifier achieves low BER for
direct-to-room-temperature data rates up to 10\,Gb/s, in the presence
of large common-mode interference in the package. Measured margins of
$\pm$8.5\% on the output bias and $\pm$28\% on AC clock amplitude
indicate that this component will not limit the performance of larger
digital designs. Design trades exist depending on rate, signal
amplitude, and the input impedance match at the receiving second-stage
amplifier. The simpler non-source-terminated design can be used where
explicit attenuation is added to the LNA inputs to improve the
impedance match, but this comes at the expense of reduced signal
power. Source termination handles the reflections typical of low-noise
LNAs, but requires twice as much voltage generated on chip as only
half of signal goes to the load. Source-termination also enables
distributed-amplifier designs that scale to higher data rates at
higher output amplitude. Direct interface to a second-stage
differential limiting amplifier is a goal for the future.

\begin{acknowledgments}

Many people contributed to the design, fabrication, and test of these
circuits. We acknowledge James Wise and Deepal Wehella Gamage for
design, Chris Kirby for fabrication, Jose Ibarrondo, Mohammed Lateef,
Yamil Huertas Morales and Jose Robinson for test support, and Anna
Herr for useful discussions.

\end{acknowledgments}

\bibliography{oa_preprint}

\end{document}